\documentclass[12pt]{article}
\usepackage{epsfig}

\newcommand{\eq}{\begin{equation}}
\newcommand{\eqx}{\end{equation}}
\newcommand{\eqn}{\begin{eqnarray}}
\newcommand{\eqnx}{\end{eqnarray}}
\newcommand{\f}[2]{\frac{#1}{#2}}

\newcommand{\qqqq}{\quad\quad\quad\quad}

\newcommand{\pd}{\partial}

\newcommand{\sdet}{\sqrt{-\det g}}

\newcommand{\Cd}{\dot{C}}
\newcommand{\Td}{\dot{T}}
\newcommand{\Phid}{\dot{\Phi}}
\newcommand{\ad}{\dot{a}}
\newcommand{\add}{\ddot{a}}
\newcommand{\Tdd}{\ddot{T}}
\newcommand{\Phidd}{\ddot{\Phi}}

\newcommand{\lm}{\lambda}
\newcommand{\al}{\alpha}
\newcommand{\bt}{\beta}
\newcommand{\arcsinh}{\mbox{\rm arcsinh}}

\title{Backreaction and the rolling tachyon -- an effective action
point of view\footnote{Preprint nos. NORDITA-HE-2003-27, KUL-TF-2003-12.}.}

\author{Yves Demasure$^{a,b}$\footnote{e-mail: {\tt demasure@nordita.dk}}\
\ and Romuald A. Janik$^{c}$\footnote{
e-mail: {\tt ufrjanik@if.uj.edu.pl}}\\ 
\\ \small
$^a$ Instituut voor Theoretische Fysica, Katholieke Universiteit Leuven,\\
\small Celestijnenlaan 200D, B-3001 Leuven, Belgium\\
\small $^b$ NORDITA,\\
\small Blegdamsvej 17, DK-2100 Copenhagen, Denmark\\
\small $^c$ Jagellonian University,\\
 \small Reymonta 4, 30-059 Krakow, Poland}

\begin{document}

\begin{titlepage}

\rightline{KUL-TF-2003-12}
\rightline{NORDITA-HE-2003-27}
\setcounter{page}{0}

\vskip 1.6cm

\begin{center}

{\LARGE Backreaction and the rolling tachyon -- an effective action
point of view  }

\vskip 1.4cm

Yves Demasure$^{a,b}$\footnote{e-mail: {\tt demasure@nordita.dk}}\
\ and Romuald A. Janik$^{c}$\footnote{
e-mail: {\tt ufrjanik@if.uj.edu.pl}}\\ 
\vskip 0.3cm
 
{\small
$^a$ Instituut voor Theoretische Fysica, Katholieke Universiteit Leuven,}\\
{\small Celestijnenlaan 200D, B-3001 Leuven, Belgium}\\
{\small $^b$ NORDITA,}\\
{\small Blegdamsvej 17, DK-2100 Copenhagen, Denmark}\\
{\small $^c$ Jagellonian University,}\\
{\small Reymonta 4, 30-059 Krakow, Poland}

\vskip 1.4cm
\end{center}

\begin{abstract}
We compute the decay of an unstable D9 brane in type
IIA string theory including backreaction effects using an effective
field theory approach. The open string tachyon on the brane is coupled
consistently to the space-time metric,
the dilaton and the RR 9-form.
The purpose of this note is to address the fate of the open 
string energy density, which remains constant if no interaction with
the closed string modes is included.
Our computations show that taking only into account
the coupling to the massless closed strings the total energy
stored in the open string sector vanishes asymptotically,
independently how small one chooses $g_s$.
We find also the large time behaviour of the fields in the Einstein
and string frames.
\end{abstract}
\end{titlepage}

\section{Introduction}

In recent years much progress has been made in the study of some
nonperturbative aspects of string theory.
The static properties of stable and unstable D-branes are by now well
understood. According to Sen's conjecture \cite{Sencon}, an unstable
brane starts rolling down the potential towards the closed string vacuum,
where there are no more perturbative open string states.
This has been studied in the `static' context from various points 
of view \cite{introEA,introSFT1,introSFT2}.

Recently some more dynamical aspects have attracted a lot of
attention, in particular the process of unstable D-brane decay in {\em real
time} starting from some initial configuration.
Since the unstable D-brane is described by some tachyon profile one is led to
study time-dependent tachyon dynamics. Such an exact time dependent solution 
of open string theory at $g_s=0$ was found by Sen \cite{RolT}. It can be
described as a free field BCFT with the insertion of an exact marginal 
operator on the boundary.
\eq
\label{e.bcft}
T_{BCFT}(t)= \tilde{\lm} \cosh \left(\f{t}{\sqrt{2}}\right)
\eqx
This boundary operator identifies the time-dependent classical tachyon 
profile living on the brane. The calculated energy density stored in the 
open strings is then naturally constant with time. This leads at the end
of the evolution to a pressureless tachyon matter.
See \cite{tds} for a selection of time-dependent solutions in different
settings.

The key question is then what are the properties of the final state of
the time evolution once we allow for a nonvanishing string
coupling $g_s \neq 0$.
Or formulated differently: is pressureless tachyon matter purely an
artefact of the `noninteracting' solution (\ref{e.bcft})?

In \cite{S,GS2,MSY} open string creation has been argued
to destabilize the tachyon matter. 
From the boundary state perspective, the time dependent tree level
couplings to the closed strings can be relatively easily computed 
\cite{RolT,SS}. Subsequently the creation of closed strings
from the decaying brane has been calculated \cite{LLM}.   
It was found, in some cases, that the total emitted energy
diverges. The problem was traced to neglecting the
backreaction of the emitted closed string excitations (gravitons
etc.).
In other words, the decaying brane emits gravitons which modify the
closed string background which in turn modifies the  evolution of the 
boundary state. However this modification 
of the closed string background and closed string self-interactions seem to be
extremely difficult if not impossible to implement in the worldsheet
perspective. 
For more recent work on string production and backreaction
see \cite{SG}.

In this note we want to study the evolution of an unstable D9 brane
in type IIA string theory. We want mainly to concentrate on the fate
of the open string energy density.

The correct theoretical framework for describing such an interacting
system would be an open-closed string field theory. However we lack
a workable concrete formalism (but see \cite{BF}) especially in the
superstring case.  
It may also be possible to embed closed strings in a purely
open string framework \cite{introSFT2}, however they may be
represented only in a rather singular form, and any description of a
closed string background in this manner seems to be completely beyond
our reach. 

For these reasons we decided to adopt an effective action approach and
couple the tachyon effective action \cite{garousi,RolT,Kutasov} 
to the low energy supergravity action for the massless closed string 
modes and to study the resulting temporal evolution. 

We note that the coupling of tachyon matter to gravity has already
been studied, however, the emphasis was on different questions than
the ones that we want to consider.
On the one hand, people studied the coupling of bosonic tachyonic
matter in 4D to (4D) general relativity and studied it as a possible
source of inflation (in the `real world') \cite{Mazumdar,Gibbons}. 
On the other hand, a more related study investigated the supergravity
solutions corresponding to SDp branes \cite{SDsugra,LP1,BW,LP2,Mc},
which were introduced in \cite{GS}. 

As stressed before, our motivation is different. We want to determine 
whether in the large $t$ limit there is still open string matter or whether
it has all been transformed into closed string modes or whether there
is some kind of intermediate solution.
As a criterion for the
disappearance of open strings we will calculate the energy density of
the tachyon matter (which is a source for the gravitational field) and
see if it vanishes in the large $t$ limit.

Moreover, from the technical point of view,  
we want to consider the full system with all the relevant
supergravity fields like the dilaton and the RR-form. In addition we
start from the tachyon below the tip of the potential (i.e. we have
static initial conditions in order for the whole evolution to come
from the decay of the unstable D-brane and not from the additional
initial kinetic energy of the tachyon), while in the case of SDp
brane solutions the opposite conditions had to be imposed \cite{LP1,LP2}.
We also choose to work in the Einstein frame in order to have clearer
notions of energy densities.

The plan of this paper is as follows. In section 2 we briefly recall
the effective action description of the rolling tachyon without
backreaction. In section 3 we derive the equations of motion for the
supergravity+tachyon system, and in section 4 we present the main
results coming from the numerical solutions and discuss the asymptotic
regime. We close the paper with a discussion.

\section{The rolling tachyon without backreaction}

An (approximate) effective action describing the dynamics of the open string
tachyon has been proposed by Sen \cite{RolT}:
\eq
\label{e.sen}
\int d^{p+1} x\, V(T) \sqrt{\det (\eta_{\mu\nu}+\pd_\mu T \pd_\nu T)}
\eqx
The particular choice of $V(T)$ \cite{coshPot}
\eq
\label{e.v}
V(T)=\f{1}{\cosh \left(\f{T}{\sqrt{2}}\right)}
\eqx
leads to the dynamics very similar to the one obtained from the exact
BCFT (\ref{e.bcft}). 
The solution to the EOM of (\ref{e.sen}) with the initial conditions
$T(0)=T_0$ and $\Td (0)=0$ is
\eq
T(t)=\sqrt{2} \, \arcsinh \left[\sinh\left( \f{T_0}{\sqrt{2}}\right)
\cosh \left( \f{t}{\sqrt{2}}\right) \right] 
\eqx   
For large $t$ one has $T(t) \sim t$. Note that this is quite
different from the BCFT profile (\ref{e.bcft}). However there may well
be some field redefinition between the two approaches. Invariant
information is encoded in the energy momentum tensor. We can thus use
the $T_{00}$ component to match the $\tilde{\lm}$ parameter of the
BCFT profile and the effective action solution:
\eq
\label{e.enmatch}
T_{00}=\f{1+\cos \left(2 \pi \tilde{\lm} \right)}{2} \equiv
\f{1}{\cosh \left( \f{T_0}{\sqrt{2}} \right)}
\eqx  
The static initial boundary conditions thus always correspond to
evolution from below the tip of the tachyon potential.

The advantage of the specific choice of (\ref{e.v}) is that the
functional $t$-dependent form of the $T_{ii}$ component is the same as
for the exact BCFT profile (note however that then the matching of
parameters is slightly different from (\ref{e.enmatch})).

\section{Coupling to supergravity fields}

We will now specialize to the the decay of an unstable D9 brane in
type IIA superstring theory. The reason for that is that we want to
have a well defined bulk closed string theory (no closed string
tachyon) and with the above spacefilling brane all the supergravity
equations reduce just to ordinary differential equations which can be
easily solved numerically.

The relevant supergravity fields will be the metric $g_{\mu\nu}$, the
dilaton $\Phi$ and the RR 9-form $C_9$. The SUGRA action for these
fields (in the Einstein frame) is
\eq
S_{SUGRA}=\f{1}{16\pi G_{10}} \int d^{10}x\, \sdet \left(R-\f{1}{2}
g^{\mu\nu}\pd_\mu \Phi \pd_\nu\Phi -\f{1}{2\cdot 10!} e^{-\f{5}{2}\Phi}
F_{10}^2 \right)
\eqx
where we used massive IIA SUGRA \cite{joep} and $F_{10}=dC_9$.

The effective action for the tachyon coming from the unstable D9 brane
is the curved space analogue of (\ref{e.sen}) with a Chern-Simons
coupling to the RR 9-form:
\eq
S_T=\f{\lm}{16\pi G_{10}} \left( -\int d^{10}x\,  e^{-\Phi} V(T) \sqrt{-\det
A}  +  f(T) \int  dT \wedge C_9 \right)
\eqx 
where
\eq
A_{\mu\nu}= g_{\mu\nu}^{str.}+\pd_\mu T \pd_\nu T=
e^{\f{1}{2} \Phi} g_{\mu\nu} +\pd_\mu T \pd_\nu T
\eqx
and
\eq
\lm =  \f{g_s}{ (2 \pi \sqrt{\alpha'})^{3}}
\eqx
For the CS coupling we take, following \cite{LP1}, $f(T)=V(T)$. We use the
potential $V(T)=1/\cosh\left(T/\sqrt{2}\right)$.

Throughout the paper we use the Einstein frame metric in order to have
a conventional interpretation for the energy density (the energy
momentum tensor is obtained using variations w.r.t. the Einstein frame
metric).

The $SO(9)$ symmetry of the unstable D9 brane decay allows us to make
the ansatz:
\eqn
ds^2&=&-dt^2 +a^2(t)((dx^1)^2+\ldots) \\
C_9&=&C(t) dx^1 \wedge \ldots dx^9 
\eqnx
and, of course, $T=T(t)$ and $\Phi=\Phi(t)$. Then the Einstein tensor
is
\eqn
G_{00} &=& 36 \f{\ad^2}{a^2} \\
G_{ii} &=& -28 \ad^2 - 8 a \add
\eqnx
and the Einstein equations are $G_{\mu\nu}=T_{\mu\nu}$. The energy
momentum tensors for the relevant fields are
\eqn
T_{00}[\Phi] = \f{1}{4} \Phid^2 &\qqqq&
T_{ii}[\Phi] = \f{1}{4} a^2 \Phid^2 \\
T_{00}[C_9] = \f{1}{4} e^{-\f{5}{2} \Phi} \Cd^2 a^{-18} &\qqqq&
T_{ii}[C_9] = -\f{1}{4} e^{-\f{5}{2} \Phi} \Cd^2 a^{-16}
\eqnx
and
\eq
T_{\mu\nu}[T] = \f{-1}{\sdet} \lm e^{-\f{1}{2}\Phi} \f{1}{2} V(T)
\sqrt{-\det A} \left( A^{-1} \right)_{\mu\nu}
\eqx
Hence
\eq
T_{00}[T] = \f{\lm}{2} e^{\f{3}{2}\Phi} \f{V(T)}{\sqrt{\Delta}} \qqqq
T_{ii}[T] = -\f{\lm}{2} e^{\f{3}{2}\Phi} V(T)\sqrt{\Delta} a^{2}
\eqx
where\footnote{Note that due to the fact that we are using the
Einstein frame, $\Delta$ is different from the one in e.g. \cite{LP1}.}
\eq
\Delta \equiv 1- e^{-\f{1}{2}\Phi} \Td^2
\eqx

In addition to the Einstein equations we have EOM for the matter
fields:
\eqn
\label{e.tach}
&& \Tdd  -\f{1}{2} \f{\dot{\Delta} \Td}{\Delta} +\Phid \Td +9
\f{\ad}{a} \Td +e^{\f{1}{2} \Phi} \f{1}{V} \f{dV}{dT} = - a^{-9} \Cd
\sqrt{\Delta} e^{-\Phi} \\
\label{e.rr}
&& \frac{d}{dt} \left( e^{-\f{5}{2}\Phi} a^{-9} \Cd \right) =\lm V(T) \Td \\
\label{e.dil}
&& \frac{d}{dt} \left(a^9 \Phid \right)= -\f{5}{4} e^{-\f{5}{2}\Phi}a^{-9}
\Cd^2 
-\lm a^9 e^{\f{3}{2}\Phi} V(T) \left(
\f{3}{2}\sqrt{\Delta} +\f{e^{-\f{1}{2} \Phi} \Td^2}{4\sqrt{\Delta}}
\right)
\eqnx

\section{Numerical results}

We solve numerically the equations (\ref{e.tach})-(\ref{e.dil}) and
the first order equation for $a(t)$:
\eq
36\f{\ad^2}{a^2}=T_{00}[\Phi] +T_{00}[C_9]+ T_{00}[T]
\eqx
The second Einstein equation ($G_{ii}=T_{ii}$) is not independent and
as a cross-check we verified numerically that it is indeed satisfied.  
We also checked explicitly that the total energy momentum 
tensor is covariantly conserved.

We choose the initial conditions $T(0)=T_0$,
$\Td(0)=\Phi(0)=\Phid(0)=C(0)=\Cd(0)=0$ and $a(0)=1$ i.e. initially at
$t=0$ we have the D9 brane in ordinary flat Minkowski space.
The initial condition for $\ad$ is not a free parameter but is
determined by the einstein equation:
\eq
\frac{\ad}{a}(0) = \frac{1}{6} \sqrt{\frac{\lm}{2 \cosh{(T_0/\sqrt{2})}}}
\eqx
Note that we always choose explicitly a positive initial
Hubble parameter.

Numerically the system of equations is difficult to solve and we had
to use high precision calculations. Nevertheless still we could not
reach asymptotic values of $t$ (e.g. $t<60$ for $T(0)=0.5$). 
The reason for the numerical
instability is the expression for the tachyon energy
\eq
\f{V(T)}{\sqrt{\Delta}}
\eqx
The numerator is exponentially suppressed, but $\Delta$ also
exponentially aproaches zero. In order to circumvent the problem we
derived an approximate expression for $\Delta$ and used it for
evolving the system to large $t$ with initial conditions
obtained at some intermediate time $t_0$ from the exact evolution. 
In this way we can reach very long
times (beyond $t=100000$ for $T(0)=0.5$). We checked that the solutions
of the asymptotic set of equations coincide almost exactly with the
exact solutions in the common domain of validity. 

Let us briefly summarize the key features of the above
simplifications.
Firstly, the equation for the RR form can be solved exactly:
\eq
e^{-\f{5}{2}\Phi} a^{-9} \Cd= \lm \int_{T_0}^{T(t)} V(T) dT
\eqx
For large $t$, since $T(t) \to \infty$ the above quantity reaches
a constant $\tilde{C}$:
\eq
\label{e.tildec}
e^{-\f{5}{2}\Phi} a^{-9} \Cd \longrightarrow \tilde{C} \equiv
\f{\lm \pi}{\sqrt{2}} -2\sqrt{2} \lm \arctan \left[ \tanh
\left( \f{T(0)}{2\sqrt{2}} \right) \right]
\eqx
Secondly, since $\Delta$
approaches exponentially $0$ we may identify (up to exponentially
supressed terms)
\eq
\label{e.tder}
\Td = e^{\f{1}{4}\Phi}
\eqx
Neglecting the terms in the tachyon EOM (\ref{e.tach}) which are
proportional to $\sqrt{\Delta}$, approximating $(1/V)dV/dT \sim
-1/\sqrt{2}$ and using (\ref{e.tder}) we obtain
\eq
\f{d}{dt} (\log \Delta)=-\sqrt{2} \Td+\f{5}{2} \Phid +18 \f{\ad}{a}
\eqx 
which yields
\eq
\sqrt{\Delta} = const^{-1} e^{-\f{T}{\sqrt{2}}} e^{\f{5}{4} \Phi} a^9
\eqx
The resulting tachyon energy density behaves asymptotically as
\eq
\label{e.ttas}
T_{00}[T] \sim \lm \cdot const \cdot a^{-9} e^{\f{1}{4}\Phi}
\eqx
The constant is fixed from the numerical solution of the
exact equations.

\subsubsection*{Tachyon energy density}

\begin{figure}[t]
\centerline{\epsfysize=40mm \epsfbox{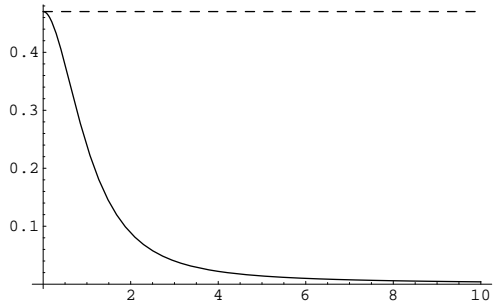}}
\caption{Tachyon energy density as a function of $t$ (for $T(0)=0.5$
and $\lm=1$). The dashed line
shows the energy density without backreaction taken into account.}
\end{figure}

\begin{figure}[t]
\centerline{\epsfysize=40mm \epsfbox{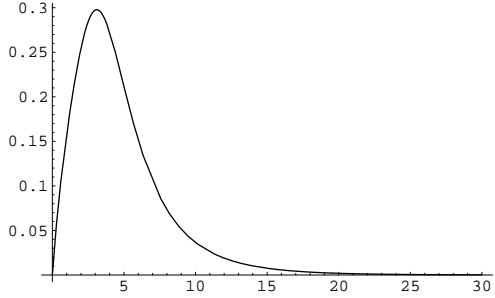}}
\caption{The coupling of the tachyon to the RR 9-form.}

\end{figure}

The main motivation for this paper was to study the influence of the
backreaction of the emitted closed string fields on the rolling
tachyon dynamics. In particular we study the behaviour of the energy
contained in the open string sector, which would remain constant
without backreaction taken into account. 
In figure 1 we plot the time
evolution of the tachyon energy density
\eq
T_{00}[T] = \f{\lm}{2} e^{\f{3}{2}\Phi} \f{V(T)}{\sqrt{\Delta}}
\eqx 
We see that it goes to zero. Moreover this is not due just to the 
vanishing of the dilaton prefactor as can be verified using the
asymptotic behaviours derived in the following section.
The same asymptotic vanishing can be seen to
hold also for the combination $\sqrt{-\det g} T_{00}[T]$. 
In figure 2 we also plot the coupling to the RR 9-form $\lm V(T) \Td$
which eventually also vanishes.

The above results indeed support the hypothesis that the whole energy
initially concentrated in the open string modes gets  transferred into
the closed string sector, no matter how small one chooses $g_s$.
However as we find below, the asymptotic
geometry is not flat static Minkowski space but rather a weakly
expanding background with nontrivial dilaton and RR 9-form fields.

\subsubsection*{Asymptotic region}

\begin{figure}[t]
{
\mbox{ }\epsfysize=40mm\hfill\epsfbox{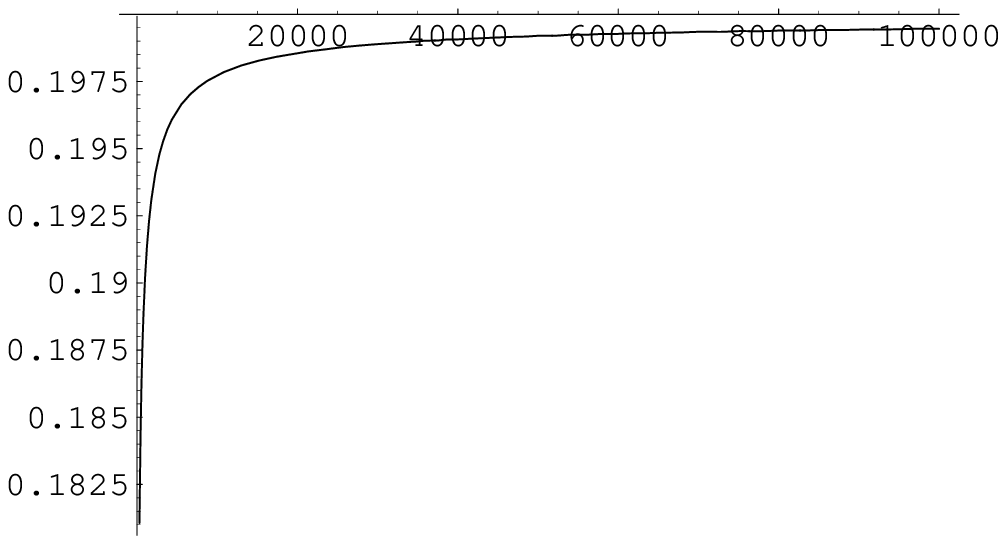}\hfill
\epsfysize=40mm\hfill\epsfbox{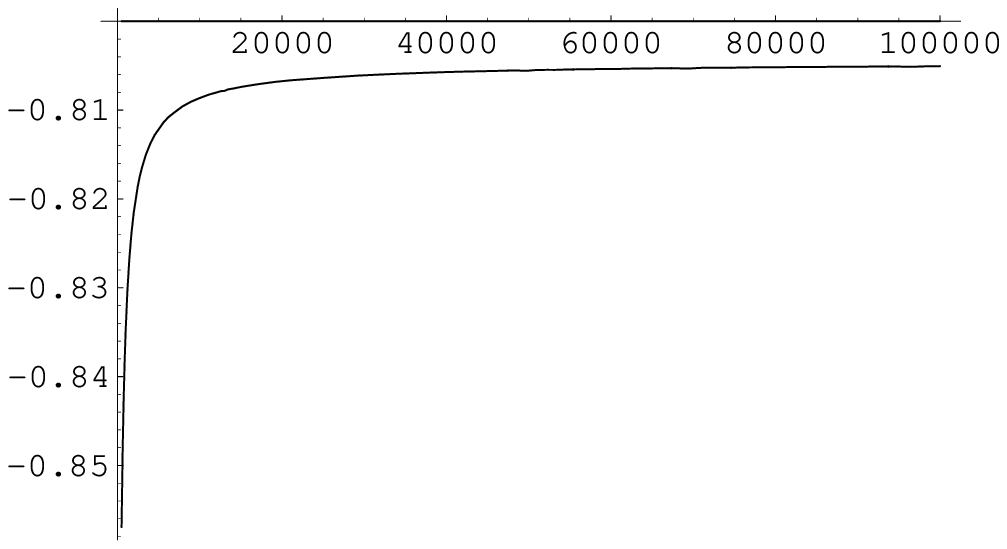}\hfill}

\caption{Verification of the asymptotic scaling behaviour of the
dilaton and the scale factor. a) $\f{\ad}{a}t$ as a function of $t$,
b) $\Phid t$ as a function of $t$.} 
\end{figure}

In the asymptotic region we obtain the set of equations
\eqn
\Phidd+9 \f{\ad}{a} \Phid +\f{5}{4} \tilde{C}^2 e^{\f{5}{2}\Phi} &=&
-\f{1}{2} T_{00}[T] \\ 
18 \f{\ad^2}{a^2}-\f{1}{8}\Phid^2 -\f{1}{8} \tilde{C}^2 e^{\f{5}{2}\Phi}
&=& \f{1}{2} T_{00}[T] 
\eqnx
where $T_{00}[T]$ is substituted by (\ref{e.ttas}) and $\tilde{C}$ is
defined in (\ref{e.tildec}).

We will now heuristically determine the asymptotic scaling dependence
of the fields. Assuming a power law dependence $e^\Phi \sim t^\al$, $a
\sim t^\bt$ and {\em requiring} that {\em all} the terms in the above
asymptotic expressions are of the same order of magnitude (i.e. $\sim
t^{-2}$) we obtain
\eqn
e^{\Phi(t)}  &\sim& t^{-\f{4}{5}} \\
a(t)       &\sim& t^{\f{1}{5}}
\eqnx
and using $\Td \sim e^{\Phi/4}$ we get
\eq
T(t) \sim t^{\f{4}{5}}
\eqx 
We verified numerically that the above scalings indeed do set in, but only
at very large times (see figure 3). Indeed the approach to
asymptotics is quite slow and due to the complexity of the equations
we were unable to quantify it further.

The RR 9-form behaves asymptotically as $C(t) \sim t^{4/5}$, which is
a direct consequence of (\ref{e.tildec}) and the above results.
With the above asymptotics, the energy momentum tensors $T_{00}[\ldots]$
behave like $1/t^2$, and although they vanish asymptotically they are
still able to drive a weak power-law expansion of the space-like
geometry. 

\subsubsection*{Asymptotic region in the string frame}

It is interesting to see how the asymptotic region looks like in
string frame since it is the string frame metric which appears in
the (closed) string sigma-model. Using the relation
$g^{string}_{\mu\nu}= e^{\Phi/2} g^{E}_{\mu\nu}$ we get for our ansatz:
\eq
ds^2_{string}=-e^{\f{1}{2}\Phi} dt^2 +e^{\f{1}{2}\Phi} a^2(t)
d\vec{x}^2 \equiv -dt'^2+ a_{string}^2(t') d\vec{x}^2
\eqx 
where we introduced natural string-frame time coordinate $t'$.
It is easy to check that the new time is related asymptotically to the
Einstein-frame coordinate through
\eq
t' \sim t^{\f{4}{5}}
\eqx
The string frame scale factor then reaches asymptotically a constant:
\eq
\label{e.astring}
a_{string}^2 = e^{\f{1}{2}\Phi} a^2(t) \sim const.
\eqx
Therefore the asymptotic metric seen by the strings is just flat
Minkowski space $-dt'^2 +d\vec{x}'^2$. 
Yet this is not the background of the classical flat space as 
the dilaton and the RR 9-form still have nontrivial $t'$ dependence:
\eq
e^{\Phi} \sim \f{1}{t'} \qqqq C \sim t'
\eqx
thus the effective string coupling constant vanishes for large times.

Note that (\ref{e.astring}) has a different behaviour than the one discussed in
\cite{LP1}. In that paper the authors found that the Einstein metric saturates
while the string frame metric collapses. 
One might think that this has to do with the different initial
condition they use:
in the SDp brane context the natural initial conditions are of the type 
$\Td (0) \neq 0$ and $T(0)=0$ which correspond to an initial energy density 
{\em above} the tip of the potential. One could thus  expect 
qualitatively that the resulting additional energy density may be enough to 
cause string-frame gravitational collapse (or stop the Einstein frame
expansion that we observe).
However we checked explicitly that these initial conditions 
$T(0)=0$ and $\Td(0) \neq 0$ lead qualitatively to the same asymptotic
behaviour that we obtained.

We also verified that if one where to continue these solutions
into the past one would encounter singular behaviour.
This however is beyond the scope
of this note as we are mainly interested in the dynamics of the time
evolution from some initial configuration and so we do not care how this
initial configuration was prepared in the first place.
See \cite{BW,LP2} for a discussion of singularity theorems in the
tachyon matter context.

\section{Discussion}

In this note we found a solution corresponding to a decaying
unstable D9 brane in type IIA string theory. The decaying
brane is described by a time dependent tachyon profile
and is coupled consistently to the graviton, the
RR 9-form and the dilaton. Note once more that we were
interested in static initial conditions with positive initial
Hubble parameter.

The resulting asymptotic spacetime dynamics is described by a 
weak power-law expanding FRW metric. 
Our computation shows that the energy density stored in the tachyonic 
open string sector is transferred completely into the closed string sector.
The large time behaviour is described by a Minkowskian string-frame metric
supplemented by a time-dependent RR9 form $C \sim t'$ and a decreasing
string coupling $e^{\Phi} \sim 1/t'$.

It would be very interesting to see how the inclusion of
massive closed strings modifies the advocated picture.
There is no a priori reason to neglect the massive closed string
states, only then we do not have an analogous effective action
description.   
However, we have shown that it is enough to include just the massless 
closed string modes to get $T_{00}[T] \to 0$. We believe that it would be very 
improbable that the inclusion of massive closed string states would  
undo this qualitative behaviour.

Note that the limit when we approach the tip of the potential is
somewhat singular. With the static initial conditions $T'(0)=0$ this
corresponds to $T(0)=0$ which leads, since the ODE's are 2nd order,
just to a constant {\em vanishing} tachyon $T(t)=0$. The precise
behaviour thus depends on the detailed form of the action for small $T$
(and possibly multiple derivative extensions) -- therefore perhaps
different treatment is needed. We leave this case as an interesting
open problem.

\bigskip

\noindent{\bf Acknowledgments} RJ was partially supported by KBN
grant~2P03B09622 (2002-2004). 
YD is supported in part by the Federal Office for Scientific, 
Technical and Cultural Affairs through the Interuniversity Attraction 
Pole P5/27 and in part by the European Community's Human Potential 
Programme under contract HPRN-CT-2000-00131 Quantum Spacetime and 
by an EC Marie Curie Training site Fellowship at Nordita, under 
contract number HPMT-CT-2000-00010.

\end{document}